# New thermodynamic constraints on internal, thermal and magnetic states of terrestrial-like Super-Earths

Mohamed Zaghoo

*Laboratory for Laser Energetics, University of Rochester, Rochester NY 14601*

## Abstract

**Ascertaining rocky exoplanets' dynamic evolution requires better understanding of key internal thermophysical processes that shaped their geological surfaces, heat fluxes, volatiles and atmospheric content. New high-pressure experiments on iron's and silicates' compressible, melting and transport properties are providing new constraints that demand reassessments of super-Earths' thermal and magnetic evolution models. We examine the interior structure, temperature distribution, thermal states and dynamo action of these planets with masses ranging from 1-10 $M_E$. We show that the shallow adiabaticity of iron-alloys and perovskite or stishovite silicates compared to their liquidus at high pressure would allow for deep basal magma oceans, and frozen iron cores in planets larger than 4 $M_E$. The presence and partitioning of MgO may alter this scenario. For the more massive planets, the dramatic reduction in liquid silicates' viscosity should ensure a vigorous convection in the lower mantle, while the rise of iron thermal conductivity under high pressures, is shown to keep the internal cores of planets more massive than 2.5 $M_E$ subadiabatic and non-convicting. This will preclude the dynamo action in the more massive super-Earths (SE). Our results could allow a new mineral physics centered classification of terrestrial-like superEarths.**

The discovery of rocky-type exoplanets transiting the circumstellar habitable zone of their host stars heralded a new era in planetary astronomy (Borucki et al. 2011; Udry et al. 2007; Vogt et al. 2010). Spectroscopic surveys reveal that some of them might harbor stable denser atmospheres rather than extended gas envelopes (Bean et al. 2010; Kreidberg et al. 2014), a condition amenable for the sustainability of oceanic water surfaces. Almost all of the rocky exoplanets detected so far receive stellar radiation that are different from Earth or have masses larger than that of our home planet. A natural starting point toward the characterization of their structure and dynamics is to establish the role of main deep processes of relevance, as a function of growing planetary mass, in shaping their surfaces, atmospheres, magmatic and magnetic properties. Of the terrestrial-like planets in the Sun's habitable zone, Earth is exceptional in possessing active plate tectonics where the lithiosphere is continuously being recycled back to the interior and the planetary surface is coupled to the internal mantle dynamics. Venus and Mars mantles exhibit a stagnant-lid mode of convection in which the lithisphere remain stagnant as a single rigid plate surface that doesn't subduct into the mantle (Schubert et al. 2001; Solomatov 1995). Earth also distinguish itself by possessing a ancient intrinsic magnetic field generated by a convicting dynamo in its molten iron core (Stevenson 2003). Venus and Mars, on the other hand, lack an active global field. Mars had an ancient dynamo, evinced by a crustal remnant magnetism (Connerney et al. 2004; Connerney et al. 1999), which persisted till ~ 4 Ga, but likely didn't survive the thermal evolution of the planet's inner core (Nimmo & Stevenson 2000). These two distinct features: magnetism and magmatism are intimately coupled through the same physical and chemical processes which drive the planetary geochemical differentiation, internal composition, core sizes and heat budgets (Buffett et al. 1996; Nimmo & Stevenson 2000). All of which depend on the behavior of their constituent materials, particularly silicates and Iron, at extreme conditions (Davies et al. 2015). The difficulty of assessing the likelihood of these conditions, and indeed the properties of SE, therefore, arise because of poor knowledge of key thermodynamic, mass and thermal and transport properties of these materials at the relevant internal conditions.

Thus far, previous studies reached conflicting conclusions regarding the vigor of internal thermal convection (Korenaga 2010; Tackley et al. 2013), the presence of an inner iron core and consequently the possibility of a self-excited dynamos in more massive SE. For example, some suggested that planets larger than 2.5 $M_E$ will be devoid of liquid inner cores, because of the steep rise of the temperature gradient of Iron's Solidus curve compared to its adiabat (Gaidos et al. 2010; Morard et al. 2011). However, core thermal evolution depends on the heat transport in the mantle, and the thus no meaningful conclusions on the thermal state of the core can be made without modeling the mantle and core as a coupled system. Others concluded that the fields generation will likely ensue in the larger planets in their early history, but the dynamo action might be short lived (Tachinami et al. 2010; Zuluaga et al. 2013). The effect of high pressure on the rheological properties of the lower mantle was studied in (Miyagoshi et al. 2013; Stamenkovic et al. 2012; Tachinami, et al. 2010), and they concluded that the increase in the subsolidus silicates viscosity at higher pressures will reduce the vigor of convection across the mantle, decrease the cooling of the core and shutting down the dynamo process in more massive planets.

The challenge with most of the aforementioned studies is that they employed geovalues for key thermodynamic or thermal transport quantities of Iron's or Silicate at exoplanet's cores or mantles (Gaidos, et al. 2010; Valencia et al. 2009). In the past few years, advances in high-pressure physics experiments, particularly those employing ramp dynamic compression tools, have addressed this challenge(Duffy et al. 2015). The experiments provided precise data on the compressibility of Iron, its gruneissen parameter (Smith et al. 2018; Wicks et al. 2018), the solidus line of $MgSio_3$ (Fratanduono et al. 2018) as well as its conductive behavior at conditions comparable to 4-5 $M_E$ (Bolis et al. 2016). Additionally, recent static high-pressure experiments revealed that liquid Iron's thermal conductivity at conditions corresponding to Earth's core mantle boundary (CMB) is substantially higher than values previously used in the geophysics literature(Gomi et al. 2013; Ohta et al. 2016). The experiments established that the thermal conductivity will further increase at higher compression. Altogether, these new thermodynamic constraints beg for a more careful assessment of the thermal and convective state of Super-Earths' cores.

Here we derive new thermodynamic data using recent experimental results on both Iron and silicates to better model the internal states for SE planets ranging from 1-10 $M_E$. We combine the state-of-the-art Equation of state (EOS) and melting experimental data with a parametric thermal evolution models to obtain new pressure, density and temperature radial profiles of these planets. We reveal that for planets more massive than 3 $M_E$, a thick layer of deep magma oceans surrounding a solid iron cores will develop. We present new theoretical data on the thermal conductivity of SE's Iron cores at extreme conditions, based on the revised estimates for Earth's values, and carefully assess the power requirements required to maintain the convective state of these cores. We show that the drastic rise in the conductive losses along the CMB will dominate the heat flux in the more massive planets, driving their cores into a subadiabatic and non-convective state. Absence substantial intrinsic heat sources, the cessation of convection will consequently shut down the dynamo action in their cores.

## Section I. EOS & Melting

For the purpose of this study, we assume that each of the terrestrial-like planets accreted into a hot early state that later underwent primordial differentiation into an iron-rich core and a primarily silicate mantle. In line with previous studies, we further assume that they are mineralogically and chemically homogenous, allowing us to approximate them by a one-dimensional radially symmetric interior model composed of two spherically concentric shells See Fig. 1. The first step in examining the internal and thermal states of our model planets is to estimate their pressure/density distributions as well as the melting relations of their constituents. This allowed us to better constrain the conditions existent in their cores and mantle, as well as the transition pressures between the layers.

Assuming that the compression is adiabatic, the equilibrium hydrostatic conditions describing the density $\rho$, pressure P, mass m and gravitational g radial profiles could be obtained by solving the continuity equations (see Appendix). Following (Valencia et al. 2006, 2007), we used the vinet EOS to calculate the radial density distribution, which provides a better description for the behavior of compressed solids at extreme pressures over its finite strain Birch–Murnaghan counterpart (Tachinami, et al. 2010). The results are plotted in Fig.2. The gradient of the adiabatic temperature profile in relation to the melting line for either the Iron or silicate determines the thermodynamic state, solid or liquid, of the layer. If the adiabatic temperature profile intercepts the iron melting line, then a solid core would exist. We determine the solidus for pure iron using the Lindemann's phenomenological law, which allows an analytical description for the melting temperature, $T_m$ as a function of pressure through the following relation. $\frac{\partial T_m}{\partial P} = 2\frac{(\gamma(P)-1/3)T_m}{K_T}$ . Whilst the adiabatic gradient is $\frac{\partial T_a}{\partial P} = \gamma\, T_a / K_S$ Here $\gamma = V\left(\frac{dP}{dE}\right)_V$ is the Gruneissen parameter, while $K_T$ is the isothermal bulk modulus. It is important to note that the Lindemann law is well motivated for simple monoatomic metals, as it describes the melting temperature as that required to displace the atoms sufficiently, in comparison to the lattice spacing, to melt the crystalline phase. As shown in Fig. 2, it is in excellent agreement with both experimental data and ab initio quantum mechanical density functional calculations for pure iron (Bouchet et al. 2013). However, as we note below, applicability for polyatomic silicate minerals at high pressures isn't well substantiated (Wolf & Jeanloz 1984).

The gruneissen parameter is a valuable thermodynamic quantity that relates the effects of density to the vibrational properties of the crystalline lattice. In most of previous studies, a constant $\gamma$, determined at the Earth's CMB, was used to derive the adiabatic and melting gradient. It is evident that the reduction in volume for simple metals at increasing densities, even if the ratio between the pressure and internal energies derivatives remain constant, will cause a concomitant decrease in $\gamma$. This was recently shown in exquisite high-pressure experiments which employed ramp-compression techniques to probe the compressibility and sound speed of Iron at pressures up to 1.4 TPa. (Smith, et al. 2018) found that $\gamma$ continued to decline as a function of pressures, reaching $\gamma \sim 0.8$ at the highest pressures documented. The density variation of $\gamma$ was determined by fitting the experimental data to the following relation $\gamma = \gamma_0(\rho_0/\rho)$, where $\gamma_0$ and $\rho_0$ are the ambient density and Gruneissen parameter respectively. For silicate minerals, the effect of increasing pressure on the bonding requires more delicate attention. A more complex chemistry dictates the structure, and the increase in coordination number in the liquid yields an increase in $\gamma$ in the region of 40-80 GPa, followed by a decline around 120 GPa to unity (Stixrude et al. 2009; Stixrude & Karki 2005). The melting line and $\gamma$ MgSio3 for perovskite was recently examined in shockwave experiments along the Hugoniot at conditions up to 250 GPa (Fratanduono, et al. 2018), it was found that the liquidus curve is better described by the Simon empirical law as $T_m = 2316[P - 20.6]^{0.177}$ (1)

The $\gamma$ for the perovskite melt was found to remain close to 1, implying that no further increase in coordination occurs in the fluid phase (Mosenfelder et al. 2009; Stixrude 2014). Although at even higher compression, $\gamma$ might decrease as is known for other high-pressure geomaterials, we take $\gamma \sim 1$ as a good approximation for the mantle conditions in SE. Integrating the equations above, with the density dependent $\gamma$ gives the solidus and adiabatic relation as a function of either density or pressure. The increase in pressure (depth variation) reduces both $\partial T_a / \partial r$ and $\partial T_m / \partial r$. However, the liquidus curve for Iron and perovskite is always steeper than the adiabat so long as $\gamma > 2/3$, see Fig. 3. Although at Earth-like conditions, the liquidus line for Fe is higher than that of the MgSio$_3$ and SiO$_2$ silicates but not that of MgO (Bolis, et al. 2016; McWilliams et al. 2012), there is a crossover that occurs at higher pressures around 400-500 GPa. Strikingly, this will mean that for terrestrial-like planets larger than 4 M$_E$, deep basal magma oceans will exist, and for the larger planets a substantial fraction of the mantle will be in the liquid state. The partitioning and fraction of MgO compared to other deep mantle silicates in these planets is currently unknown, but assuming a composition similar to that of Earth, small fraction of solid MgO would still allow for these basal oceans. At these planets, MgO might eventually gravitationally settle toward their core mantle boundaries providing some additional heat source. The arrows on Fig.3 indicates the onset of the melt of the mantle. As is shown for 5 M$_E$ SE, a magma ocean of ~1000 Km exists, while for a 10 M$_E$, this ocean extends to more 3000 Km, almost 25% of the planetary radius.

Such result stands in stark contrast to previous studies which assumed that the state of the mantle is subsolidus (Gaidos, et al. 2010; Korenaga 2010; Tachinami, et al. 2010; Valencia, et al. 2006, 2007; Zuluaga, et al. 2013). The variance between the current results and previous work is due to the fact that earlier studies used the Lindemann relation to describe the liquidus line in silicates. As we show, this empirical relation doesn't hold for silicates melt at higher pressures. Our inferred magma oceans will likely persist over the planetary timescales owing to the low thermal conductivity of molten silicates. Even with the metallization of silicate melts at higher pressures, which will result in an increase of its conductivity, the experimental shockwave data shows that these silicates still exhibit poor conductivity in their metallic regime. We further examine the effect of this result on the convective state and viscosity below.

Light alloying impurities in the cores of rocky planets will cause a depression in the pure Fe melting line. Assuming that any light constituent will reside in the liquid fluid phase of the cores, one can calculate the reduction in the melting temperature (See Appendix) by

$$\Delta T_{Fe-alloy} = -T_m \sum_i \ln 1 - \chi_{i,l} \quad (2)$$

The exact composition and concentrations of light elements in SE cores is uncertain. The initial abundance of these light impurities will clearly vary between exoplanetary systems and between individual exoplanets depending on the planet's location in the accretion zone as well as the chemical evolution of the galactic or planet-forming disks. We use a variety of concentration of S, O, Si, C to explore the range of the iron alloy eutectic temperature. The exact values are shown in Table 2. The range of $\Delta T$ for the concentration and constituents explored is 800-1500 K, and the resulting solidus for the iron alloy is plotted in Fig. 3.

## Section II. Thermal profile & Viscosity

To calculate the thermal structure of these rocky exoplanets, we rely on the widely used thermal boundary layer (BLT) model for mantle convection(Stevenson et al. 1983). The model is based with a well-substantiated assumption that terrestrial-like planetary mantles, in their liquidus or solidus state, are undergoing vigorous convection. It follows that there are thermal boundary layers at the top and bottom of the mantle. At these boundaries, heat is transported by conduction through a thermal layer of thickness $\delta$, and the temperature vary linearly with depth with a temperature jump $\Delta T$ across the boundary layer. The conductive heat flux across the layers is $F = k\,\Delta T / \Delta r$, where k is the thermal conductivity. Below and above these boundaries, the mantle convective temperature profile is described by the adiabatic gradient

$$\frac{\partial T_a}{\partial r} = \frac{\gamma(r)\rho(r)g(r)T}{K_s} \quad (3)$$

The temperature jump across the boundaries are related to the thickness of these boundary layers and the viscosity via the Rayleigh number through $R_a = \frac{\rho g \alpha \Delta T D^3}{\eta \kappa}$. Here $\alpha$ is the volumetric coefficient of thermal expansion, $\kappa$ is the average thermal diffusivity in the mantle and D is the thickness of the convective region. It follows from Fig. 3 that a considerable fraction of the deep basal mantle in the more massive planets will be liquid. Two key consequences follow from this result, first a dramatic reduction in the viscosity by 20-23 order of magnitude. Second, an increase in the thermal diffusivity, $\kappa$, of the mantle conditions. $\kappa = \lambda / \rho C_p$, where $\lambda$ is the thermal conductivity where $C_p$ is the heat capacity. Accurate determination of this value is possible from shockwave experiments, but is beyond the scope of this paper. However, it is evident that the increase in the thermal conductivity will be offsetted by the increase in the heat capacity and the density at the corresponding conditions.

Assuming that the heat flux at the core mantle boundary determines the thickness of the boundary layer $\delta = a\frac{D}{2}\left(\frac{R_a}{R_{ac}}\right)^{\beta}$ (4) With $\beta = -1/4$ and a is the order of unity (Tachinami, et al. 2010). In Earth, the exact contributions from radiogenic decay in the mantle, secular cooling or conductive heat flow from the cores is still uncertain, and thus $\delta_{CMB}$, and its associated temperature jump, $\Delta T_{CMB}$, remains ill-defined. If the scaling above holds for rocky planets different than Earth, then assuming a similar rotational period similar to that of Earth, and for a constant $R_{ac}$, $\delta \propto \eta^{1/4}$. This means that the precipitous decline in viscosity in the lower mantle of rocky planets larger than 3 $M_E$ will further thin out the lower boundary across their CMBs. This reduction could be further exacerbated by ejection of thermals and plumes as a result of buoyancy instability arising in the low viscous magma oceans. The thinning will reduce the temperature contrast and might erase it all together in the more massive SE. To calculate the entire thermal radial profile, we followed the procedure introduced in (Tachinami, et al. 2010; Yukutake 2000), where a surface boundary layer with a fixed 1500 K jump separates the surface from the upper mantle. The surface temperature was fixed at 300 K, a condition required to support liquid water. An adiabat is drawn from this boundary layer to the CMB using Eq. 12, where the entire mantle is taken to be fully convecting. In lieu with previous studies, for all of our nominal terrestrial planets, we assumed a mass ratio between the core and mantle of 3:7. $\Delta T_{CMB}$ initially increase from 1 to 2 ME but is then reduced for the larger planets. In (Sotin et al. 2007; Valencia, et al. 2006), they used a parametrized convection model that assumed $\Delta T_{CMB}$ is mass independent, while (Tackley, et al. 2013) found that $\Delta T_{CMB} \sim 0$ in SE, assuming no heating from the cores. Here, we used both a fixed $\Delta T_{CMB} \sim 1200$ K and a reduced one for more massive SE. The core temperature profile is also calculated using the Fe adiabat set at the CMB temperature.

Our thermal structure as a function of increasing planetary mass is plotted in Fig. 4, and could be compared to the Fe-alloy liquidus curve. A particularly significant result shown is that only for planets with masses of 1 to ~4 $M_E$, does the Fe adiabat cross the liquidus line. This result holds for the range of $\Delta T_{CMB}$ values explored here. Indeed different $\Delta T_{CMB}$ changes the initial starting point of the Fe adiabat and subsequently the final temperature at the SE planetary cores, it doesn't alter the key conclusion that the cores of the more massive planets will be frozen out, absent additional heat sources. For SE with 1-4 $M_E$, an inner solid core and an outer fluid core shall exist. As these planets cool, so will their adiabats, and the intersection point between these adiabats and the liquidus curves will move to shallower depths as the temperature of the inner core boundaries decrease. The resultant growth of these planetary cores will drive the expulsion of light alloying elements and provide a heat source through the released latent heat of crystallization. This heat flow, in additional to internal radiogenic decay, will heat the mantle which will be cooling radiatively through the planetary surface. Our calculations of the melting and adiabatic gradient of the cores SE, thus, affords insight into their thermal states. Of a specific importance is whether these cores at conditions corresponding to more than 1 $M_E$ are still undergoing the same thermal convection their mantles are. This question is all the more relevant with recent theoretical and experimental results strongly suggesting a higher Fe thermal conductivity at geological conditions of the Earth CMB than previously assumed. These results have brought the Earth internal energy budget under considerable scrutiny prompting revisions to previous estimates on the cooling history of Earth's inner core, as well as the age of our geodynamo. On this vein, it is important to highlight that almost all previous studies on SE just assumed that the mantle convection will cool the cores in massive rocky exoplanets, as is the case in Earth, and the resulting buoyancy forces will keep their iron cores convective. Below we assess this question using an analytical model that accounts for the changes in the thermal conductivity, thermal expansivity, adiabatic bulk modulus under increased compression.

## Section III. Thermal states of the cores

The stability of the core convection necessitates a total heat flow across the CMB, $Q_{CMB}$ larger than the core adiabatic conductive loss, $Q_a$

$$Q_a = k_c \frac{\partial T_a}{\partial r}\bigg|_{CMB} = k_c \gamma \rho T_{\text{CMB}}/K_S \quad (5)$$

If $Q_{CMB} > Q_a$, the core is superadiabatic and convective while for $Q_a < Q_{CMB}$, the core becomes subadiabatic. This condition represents the minimal requirement for the persistence of thermal convection which could otherwise be depressed by a stable chemical stratification. Another way to express this is that the core's temperature gradient has to exceed its adiabatic one. We don't attempt to calculate the total heat flows in these SE cores. This effort was explored in previous work (Tachinami, et al. 2010; Zuluaga, et al. 2013), albeit the use of geovalues for most of the relevant thermodynamic parameters, like $\gamma$, $k_c$, $K_S$. Here, we take a simpler analytical approach that permits a calculation of the temperature fields inside planetary cores by relating it to the concomitant outward growth of an inner core by solidification. As noted, if the temperature profile exceeds the adiabatic gradient, then their cores are superadiabatic and thermally convecting. We rely on the model introduced by (Buffett, et al. 1996; Lister & Buffett 1995) where the rate of cooling the planet's inner core can be used to obtain a solution for this core's radial temperature profile. The Buffet et al. model is based on the conservation of energy, mass and momentum which couple the size of the inner core as well as its thermal evolution to the mass transport of light elements, gravitational and latent heat release. As we remarked above, the cooling process of the core drives the release of gravitational energy, which is then converted to heat by both ohmic dissipation and thermal convection. Additional gravitational energy is released as the result of the expulsion of light alloying impurities from the solid to the outer core, a process that lead into compositional buoyancy. This cooling, and the resultant inner core growth, is determined by the difference between heat flow at the CMB, $Q_{CMB}$, and the radioactive heating in the mantle $Q_{rad}$ as (Eq.1 of (Buffett 2009)

$$\begin{aligned} Q_{tot} &= Q_{CMB} - Q_{rad} \\ &= H[2\chi \\ &+ 3(G+L)\chi^2]\frac{\partial \chi}{\partial t} \quad (6) \end{aligned}$$

H is the heat required to cool the whole core to its solidification temperature, G and L are dimensionless parameter which characterize both the gravitational energy and latent heat of solidification release, where $\chi$ denotes a dimensionless time dependent inner core radius normalized by the CMB depth $\chi = R_c(t)/R_{cmb}$.

H could be expressed as

$$\text{H} = 4\pi \left(\frac{1}{3} - \frac{\emptyset}{5}\right) \rho_0 C_P \; \theta \; R_{CMB}^5 \left[\frac{\partial T_L}{\partial P} - \frac{\partial T_a}{\partial P}\right] \quad (7)$$

$$\emptyset = \theta \; R_{CMB}^2 \; \gamma_0/K_0$$
$$\theta = 2\pi \; G \; \rho_0^2/3$$

Because the latent and gravitational heat release is proportional to the size of the inner core, solving for $\chi$ in the above Eq. 17 gives

$$\chi = \sqrt{t(Q_{tot}/H)} \quad (8)$$

As shown in (Buffett, et al. 1996), one can express the temperature radial profile T(r,t) as a function of the core central temperature and the cooling rate as

$$T(r,t) = T(0,t) - \Delta T(r)(r/R_c)^2 \quad (9)$$
with 0<r<$R_c$, where r=0 is the center of the planet

$\Delta T$ is the temperature drop across the core radius and is related to the total heat conducted across the CMB by the heat equation by $\frac{\partial T}{\partial t} = -6 \; \kappa \; \Delta T/R_{cmb} \sqrt{\frac{(Q_{tot})}{H}}$. We are primarily interested in this $\Delta T$ when compared to the temperature drop across the core's adiabat $\Delta T_a$. Since T(r=0,0) is the melting temperature of Fe at the central planetary pressure $T_m(P = P_0)$, one can readily express the temperature profile in terms of the melting profile as

$$\Delta T = \Delta T_m \left(1 + \frac{6\kappa}{R_{cmb}\sqrt{\frac{Q_{tot}}{H}}}\right) \quad (10)$$

Thermal convection ensues only when $\Delta T > \Delta T_a$.

We have calculated the $\Delta T$ as a function of planetary mass in order to ascertain the possibility of thermal convection in their Fe cores. For all our input parameters, we took the effect of increasing pressure and temperature to more realistically mimic the heat transport in the SE more massive interiors. $R_{cmb}$, $R_{center}$ were taken from our EOS plots. The difference between the melting and adiabatic pressure gradient $\frac{\partial T_L}{\partial P} - \frac{\partial T_a}{\partial P}$ and the ratio between the adiabatic and melting temperature difference across the cores for the different SE was readily calculated from Fig. 2. The rest of thermodynamic parameters are shown in Table 3.

As is clear from Eq. 21, $\Delta T$ depends crucially on the thermal conductivity of Fe. This is one of the most critical parameters in Earth's thermal and magnetic history. Recently a multitude of sophisticated theoretical ab initio calculations have pointed out that $k_c$ is in the order of 100 W/m.K instead of the previously used value in the geological literature ~ 40 W/m.K (Koker et al. 2012; Pozzo et al. 2012). This was later experimentally confirmed in static compression studies combining electrical resistivity measurements with laser heating techniques. The experiments reported $\rho_{Fe\;(CMB)} = 1.5 \; \mu ohm$ and used the Wiedemann Franz law to calculate the $k_{CMB}$ at measured temperatures corresponding to that existent in the Earth's CMB (Gomi, et al. 2013; Ohta, et al. 2016). Unsurprisingly, the data confirmed that the resistivity decreases at increasing pressures for the same temperatures. This is well understood for simple metals, considering the Drude-Botlzmann formula $\rho = m_e/n_{eff}e^2\tau$, where $n_{eff}$ is the number density of the effective conduction carriers, while $\tau$ is the scattering time (Bardeen 1940). In the Debye theory for crystalline metals, $\tau$ is proportional to both $T/\Theta^2$, with $\Theta$ being the Debye characteristic temperature, and the density of states at the top of the Fermi surface. For transition metals, the reduction in the interatomic spacing at higher densities will reduce the vibrational amplitude and increase in the density of states resulting in a concomitant increase in $\tau$ and a reduction in $\rho$. In liquid metals, the Ziman formalism for weak scattering have a similar dependence on the Fermi wave vector, and an increase in density result in lower resistivity (Zaghoo & Collins 2018; Ziman 1960).

To combine the effects of high temperatures and pressures we have used the Matthiessen rule to determine the electrical resistivity of Iron, where

$\sigma = \frac{1}{\rho_{tot}} = 1/\rho_{bloch-gruneisen} + 1/$

$\rho_{sat(ioffe-regel)}$,

see ((Gomi, et al. 2013; Ohta, et al. 2016). Here,

$\rho_{bloch-gruneisen} = D(V)(T/$

$\Theta(V))^x \int_0^{\Theta/T} \frac{z^x}{(\exp(z)-1)(1-\exp(-z))} dz$  (11)

$$\rho_{sat(ioffe-regel)} = m_e \hbar (4\pi^2)^{1/3} / 3n^{1/3} e^2$$

We assumed a constant x=1 for pressures higher than 200 GPa and obtained the volume reduction from our EOS. The thermal conductivity was obtained from the Wiedemann Franz law which states which relates the thermal to electron transport by $k = LT\sigma$, where L is a proportionality constant $2.44 \times 10^{-8}$. The increase in k at SE internal core conditions is more dramatic than $\sigma$, because of the additional linear dependence of temperatures which varies from ~3000 K at Earth's CMB to ~10000 K at a 10 $M_E$ planet's CMB. Our determined k values are in good agreement with ab initio calculations to their highest reported pressures, up to 350 GPa. Table 3 shows the calculated thermodynamic and transport input used to compute the temperature gradient in SE. We have assumed a similar radiogenic heat decay in the more massive planet's mantle to that in Earth's, and that the change in the entropy for the latent heat release doesn't feature a pressure dependence.

In Fig. 5 we plot the temperature profile for different SE planets as a function of the heat extracted across their respective CMBs. The solid lines denote their adiabatic temperature drop across their cores $\Delta T_a$. The core's temperature profile grows primarily shallower for the increasingly massive SE as a result of the enhanced compression effects on the thermal transport. In Earth's, the temperature gradient exceeds the adiabatic one for $Q_{CMB}$ >13 TW, stabilizing thermal convection. For heat loss less than this, any possible convection would be driven compositionally. We note that this value is in remarkable agreement with recent estimates for Earth's thermal evolution. In those studies, a different approach than the one currently used here was employed to determine the geodynamo entropy as a function of heat loss across the CMB. Positive entropy is needed to sustain the dynamo action. For 2 $M_E$ SE, thermal convection becomes possible for $Q_{CMB}$ > 50 TW. Such high values could be possible considering a different concentration of radiogenic elements arising from different accretion scenario, resulting in higher radiogenic heating. More markedly, the temperature profile grows remarkably shallow for the more massive SE, in so much it would not meet the adiabatic gradient for any realistic $Q_{CMB}$ values. This means that the cores of SE larger than ~2.5 $M_E$ will remain subadiabatic and non-thermally convicting. Taken with the above conclusion in Section II regarding the liquidus line of Iron-alloy, only rocky planets cores with 1-~ 4 $M_E$, could support a dynamo action.

### Discussions

Throughout this work, we have ignored any possible structural changes in the solid mantle before the onset of melting or in the crystalline ultra-compressed Iron above 1.4 TPa. For prototypical mineral silicates, perovskite $MgSiO_3$ or $SiO_2$, experimental data up to 400 GPa appears supportive of this conclusion. In Iron, high pressure ramp compression data up to 1.4 TPa doesn't feature a discontinuity in the P-V space which is characteristic of a phase transition. Nonetheless, even if such phases do exist, they have little consequence on the thermal structure derived here and the conclusions that followed. Below we summarize the key features that distinguish our current internal & thermal models from those previously reported:

1- A deep basal liquid mantle will develop in SE more massive than 4 $M_E$. These magma oceans will grow thicker as a function of increasing planetary mass, reaching 25% of the planetary radius in a 10 $M_E$ SE. They will also likely persist over the planetary time scale considering the steep rise of the liquidus line relative to the mantle adiabat.

2- SuperEarth planets larger than 4 $M_E$ will be devoid of fluid cores as Iron assume a crystalline phase at the corresponding planetary internal conditions. This will deprive them of major heat sources (latent heat of crystallization and gravitational energy release arising because of light elements expulsion).

3- As for the thermal structure, the rise in the thermal conductivity, due to increased compression and temperature, will dramatically increase the amount of heat loss across the cores of the more massive SE. This will mean that the cores of rocky planets larger than 2.5 $M_E$ will not sustain thermal convection over their geological timescales, but will rather cool to a conductive subadaiabtic state. This result stands independent of point 2 outlined above, and further precludes the possibility of long-lived magnetic protection for these SE. In another word, even if the planetary adiabats rises insomuch as it exceeds the Iron liquidus line in massive SE, these cores will likely remain conductive.

A diagram summarizing these finding is shown in Fig. 6. It is important to note that the value of 4 $M_E$ isn't intrinsically significant. However, it emerges as a result of the effects of increased compression and temperature on the internal thermodynamic states of SE. The formation of thick dense magma ocean will have significant consequences on the thermal evolution and global chemical differentiation of the massive SE. In Earth, partial melting of the mantle through upwelling and decompression drives the chemical segregation of silicates and volatiles cycle and ultimately give rise to our oceanic and continental crust. Although the details of this process, in particular the timescale of crystallization and the depth origin at which it ensues, remains poorly constrained, the melt viscosity is the key property dictating these processes. In the more massive SE, the highly turbulent convection, inferred from the colossal Rayleigh number in these oceans, could effectively prevent chemical differentiation by suppressing crystal settling (Tonks & Melosh 1990). The partition coefficients of trace & noble elements at these extreme conditions are not well understood. However, these coefficients would dictate whether these elements partition into the liquid or the solid phases, ultimately shaping the chemical evolution of the mantle as well as the global budget of volatiles in the crust (Monteux et al. 2016). If these trace elements preferentially fractionalize into the deep liquid, it would enhance large scale cumulate overturns (Elkins-Tanton et al. 2005), whereas if they migrate toward the upper solid mantle, they would catalyze volcanic activity (Moyen & Herve 2012). In this respect, our results should invite a revision to previous estimates on the possibilities of volcanism and plate tectonics in the more massive SE.

The bearing of the current results on the persistence of a self-excited dynamo action in SE is noteworthy. Planets only up to ~2.5 $M_E$, provided sufficient intrinsic heat sources, would possess thermally convicting liquid iron cores that would sustain an intrinsic magnetic field similar to that of Earth. For the more massive rocky planets, the cooling is shown to be insufficient to maintain thermal convection in the entire core, and a thick conducting layer would develop below CMB. Absent a growing core in SE larger than 4 $M_E$, this layer will be thermally stratified. The deep magma oceans raise an intriguing prospect of a possibly convicting and conducting silicates shell that, in principle, is capable of sustaining a dynamo. This point merits further investigation, and is partially motivated by shockwave experiments evincing a metallization for the silicates along their hugoniont above 1 TPa. It should be cautioned, however, that the same experiments show that mineral silicates at 6-8 Mbar and 7000-9000 K is, at best, a poor semiconductor, with an electrical conductivity of only few tens S/cm (Bolis, et al. 2016). This value is remarkably low especially when contrasted with liquid iron which possess a conductivity that is 1000 times higher. If these poorly conducting oceans cannot produce a dynamo action, the presence of a more conducting iron shell beneath could severely attenuate the field because of an electromagnetic skin depth effect similar to the one proposed for Saturn (Stevenson 1982).

**Conclusions**

We have studied the internal and thermal states of rocky planets with masses ranging from 1-10 $M_E$. The recent availability of high-pressure experimental data enabled us to quantitatively better characterize the melting, density, temperature structure and magnetic activity of these planets. Our results suggest a dramatically different picture for the more massive SE than the one previously discussed. Planets with 1-4 $M_E$, share similar overall structural characteristics to that of Earth, in terms of a solid convecting mantle surrounding an outer liquid iron core and a growing inner solid one. For SE planets more massive than ~2.5 $M_E$, their cores would cool to a non-convective state. These planets are likely to sustain an active dynamo capable of providing magnetic protection for their both their atmospheres and lithospheres. However, for planets larger than 4 $M_E$, their internal structure will exhibit an inverted state of solid subadiabatic iron core surrounded by a thick convicting liquid magma ocean and an upper solid mantle. Such planets could feature markedly different chemical, thermal evolution, volcanic activity and atmospheric properties. The lack of active magnetic protection may have implications toward their potential habitability, especially if it is extended across geological timescales. Our results show that mineral physics provides an important approach toward the classification of Earth-like worlds, and could lend support to the recently proposed concept of "super habitability", employed to describe terrestrial-like planets with enhanced characteristics amenable to their habitability (Heller & Armstrong 2014).

EOS parameters for Mantle and core layers used in SE radial profile calculation shown in Fig. 2.

| Material | $\rho$ (kg m$^{-3}$) | $K_0$ (GPa) | $K_0'$ (GPa) |
|---|---|---|---|
| olivine | 3347 | 126.8 | 4.27 |
| Wd+rw | 3644 | 174.5 | 4.27 |
| Pv+fmw | 4152 | 223.6 | 4.274 |
| Ppv+fmw | 4200 | 223.6 | 4.52 |
| Fe | 8300 | 177 | 5.6 |
| FeS | 5330 | 126 | 4.8 |

Calculation of the melting temperature of several Iron-alloys resulting from different light alloying impurities concentrations. The different concentrations account for a variety of possible compositions arising from different accretion scenarios.

| Impurity | $S$ | $Si$ | $O$ | $\Delta T/T$ | $\Delta T_m$ (K) |
|---|---|---|---|---|---|
| Wt% | 11 | | | 0.196 | -1174 |
| Wt % | 3 | 4 | 1 | 0.159 | -970 |
| Wt % | 2 | 0.5 | 4 | 0.175 | -1070 |
| Wt % | 4 | 1.2 | 1.2 | 0.143 | -871 |
| Wt % | 6 | 4 | 2 | 0.234 | -1402 |

| Wt % | 15 | | | 0.27 | -1615 |
|------|----|---|---|------|-------|
| Wt % | 9 | 5 | 1 | 0.27 | -1625 |
| Wt % | 5 | 9 | 0 | 0.25 | -1528 |

Thermodynamic parameters for the core energetics in SE planets used in the calculation of the thermal states of these planetary interiors. The Gruneissen parameters are extracted from Smith et al. 2018, pressures are taken from radial profiles in Fig. 2, while the gradient of the temperature dT/dr were taken from the thermal profiles plotted in Fig. 3. Calculation of the thermal conductivity was done using the electrical conductivity determined above assuming the Wiedemann Franz law.

| $M_E$ | $\gamma_{Fe-cmb}$ | $P_{cmb}$ GPa | $\gamma_{Fe-core}$ | $K_s$ GPa | $P_{core}$ GPa | $k_{e-Fe}$ W/mK | $-k_e\,\delta T/\delta r$ mW/m² |
|-------|-------------------|---------------|--------------------|-----------|----------------|------------------|-------------------------------|
| 1 | 1.36 | 136 | 1.25 | 1300 | 360 | 110 | 13.5 |
| 2 | 1.26 | 343 | 1.1 | 1850 | 757 | 272 | 19.2 |
| 5 | 1.13 | 676 | 0.8 | 2847 | 2447 | 391 | 44.5 |
| 10 | 0.93 | 1354 | 0.76 | 6834 | 4732 | 579 | 55 |

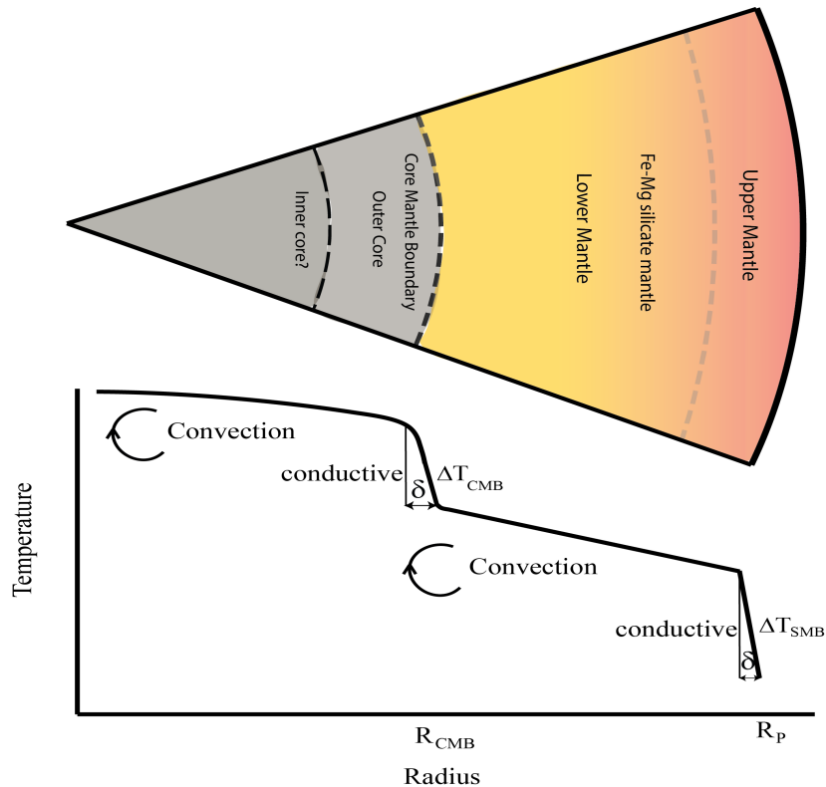

Figure 1. Top panel shows a schematic of a terrestrial-like planet's interior composed of Silicate mantle and an Iron core. The different boundaries demarcating the transition between the upper/lower mantle and the outer and inner core are also shown.

Bottom panel shows a schematic of an internal radial thermal profile for the same rocky planets assuming the thermal-boundary layer model. The planet's interior is mostly convective (temperatures follows an adiabat) except at the surface-mantle boundary and the core mantle boundary where a temperature jump occurs. Note that consistent with Earth-thermal structure, the upper-lower mantle transition doesn't constitute a boundary for convection.

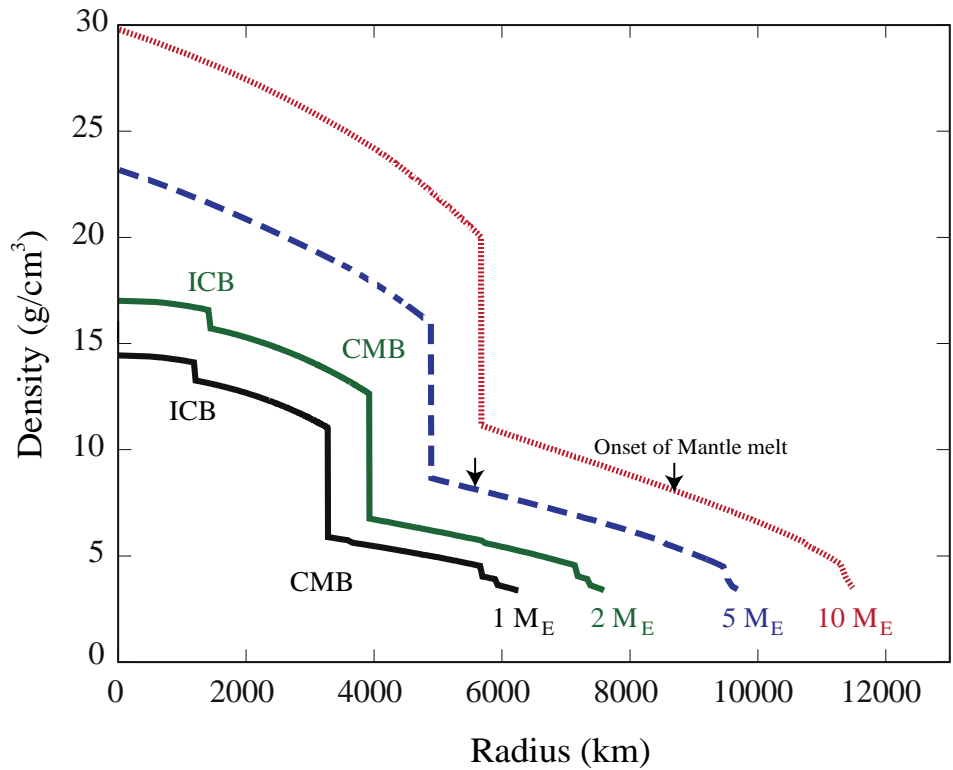

Figure 2. Radial EOS profile of the SE planets ranging from 1-10 ME. The thermodynamic parameters used to calculate the radial density profiles are shown in Table1. Planets more massive than 4 $M_E$ distinguish themselves from less massive SE in two key features: 1- the absence of a liquid outer core as is evident in the lack of Inner core boundary in the radial density plots.2- The presence of deep basal magma oceans marked with the arrows as the melt line crosses into the planetary adiabat (see Fig.3). We haven't accounted for a small density jump from the melt in the mantle silicates, although it is calculated for the Early Earth to be in the order of ~0.1-0.2 g/cc.

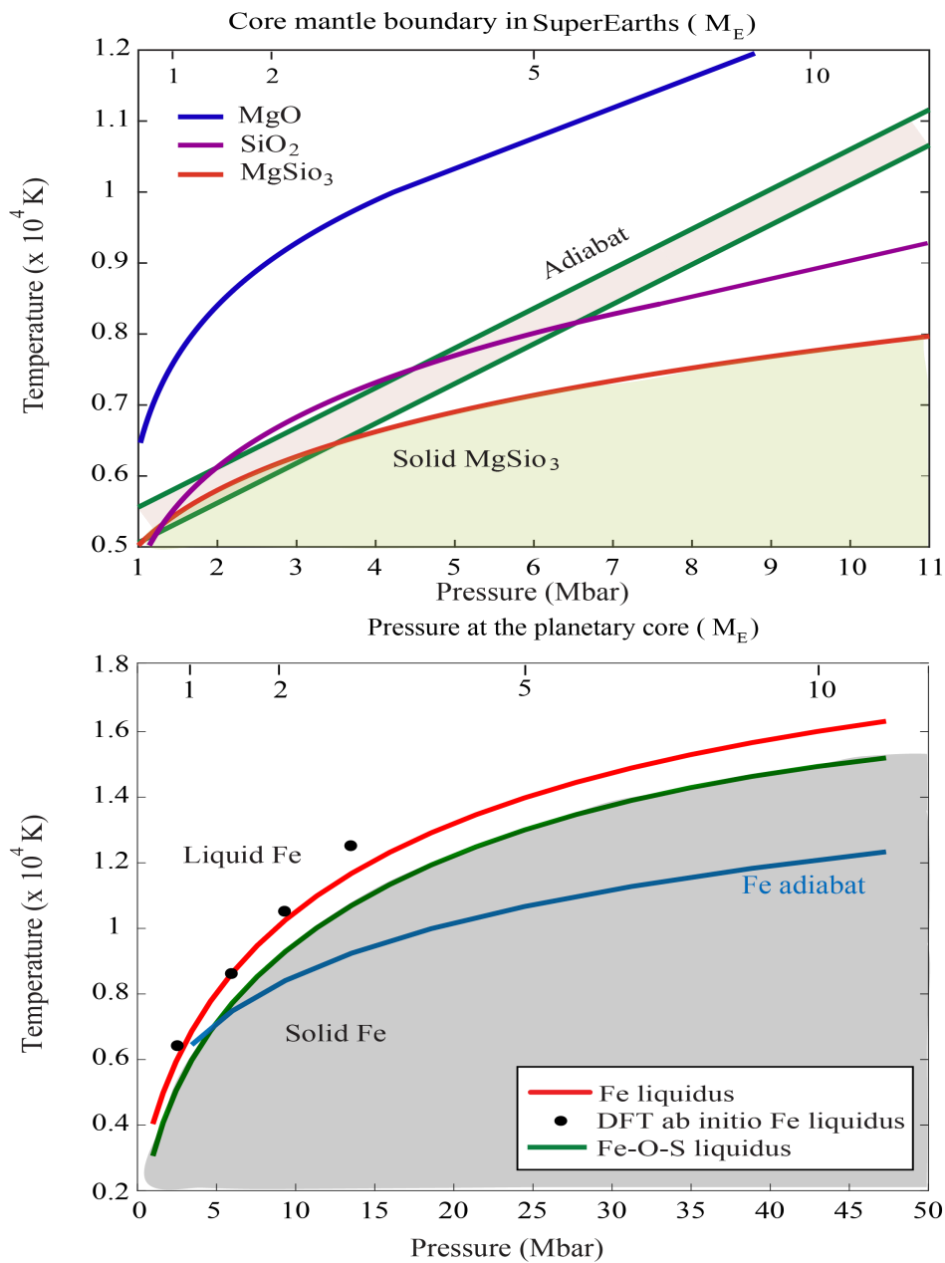

Figure 3. Top panel: The phase diagram of mantle silicates: Perovskite MgSiO₃, primary constituent of the mantle, SiO₂, MgO showing their liquidus in comparison to planetary adiabat lines. Two adiabats are shown from two different initial temperatures 5000 and 5500 K. The adiabat becomes steeper than the liquidus above 4-5 Mbar for both SiO₂ and MgSiO₃.

Bottom Panel shows the phase diagram of Iron. The adiabat of pure Iron and the Liquidus of Iron-alloy is also shown, and were calculated using formalism in the main text. The reduction in the liquidus of Iron due to impurities ranges from 1000-1500 K see Table 2. Also shown is the pure Fe Density functional theory calculation of the melt line, which is in excellent agreement with the liquidus we calculate here using the Lindemann relation.

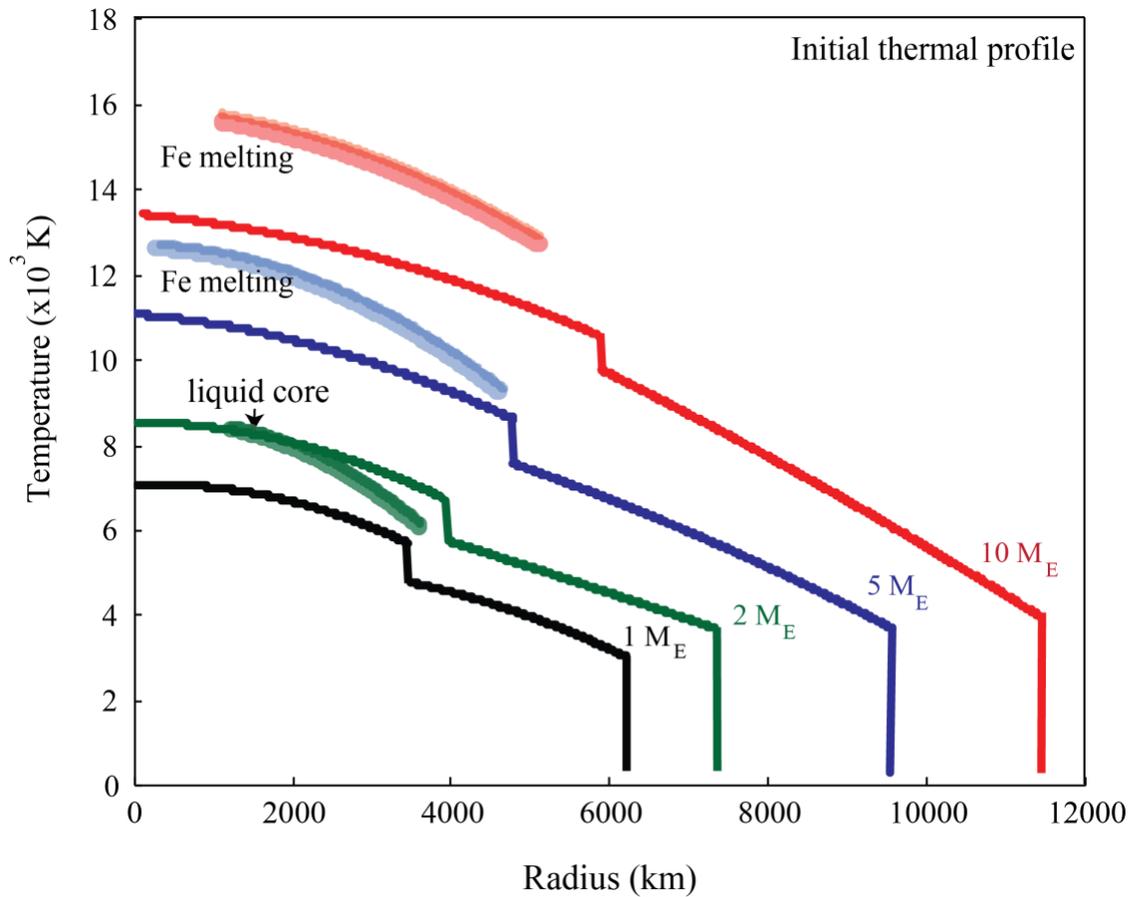

Figure 4. Radial thermal structure profiles of rocky SE planets ranging from 1-10 $M_E$ calculated within the thermal boundary layer model. Also, shown is the pure Fe-liquidus lines. The shaded area represents the reduction in the melting line due to added impurities. The intersection of the thermal profile with the liquidus denote the onset of iron crystallization, the inner core boundary. For planets 4-10 ME, the liquidus doesn't cross the planetary thermal profile, meaning that these planets will likely lack an iron fluid core.

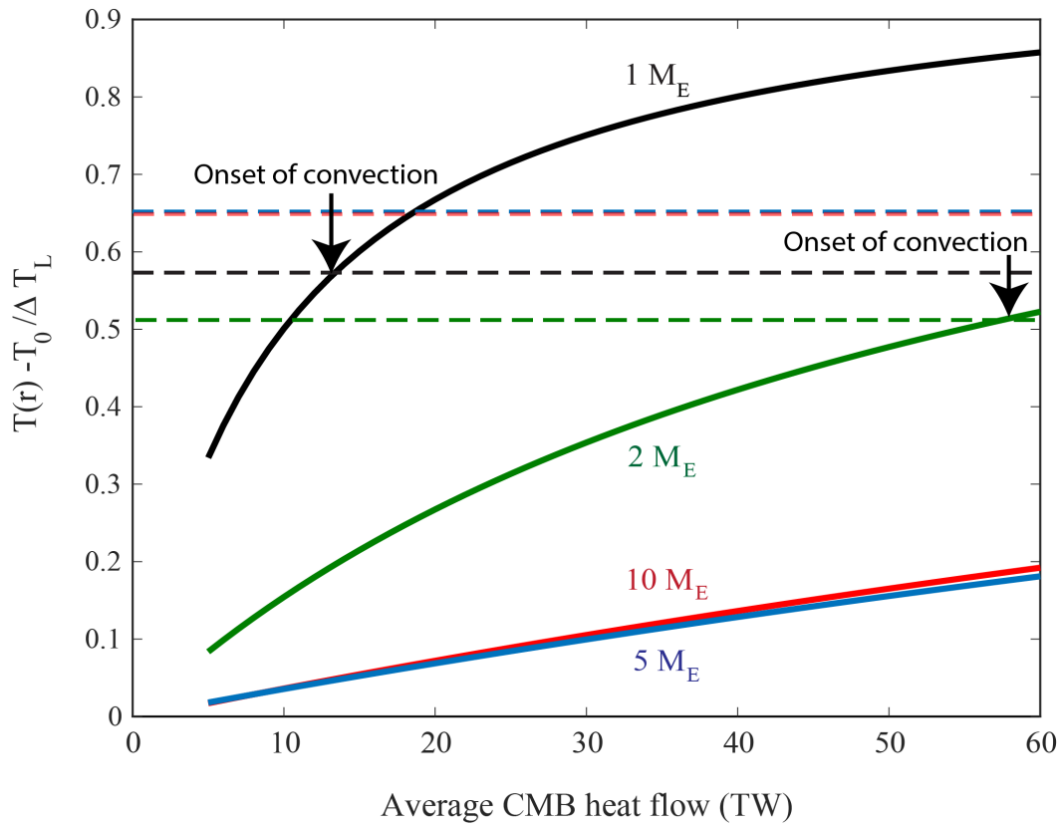

Figure 5. The difference in the temperature across the planetary iron core normalized by the difference in the melt temperature across the same conditions plotted as a function of average conducted CMB heat flow. The vertical lines show the adiabatic values $\Delta T_a/\Delta T_L$ calculated for different SE planets from Fig. 4. Thermal convection is possible only when $\Delta T > \Delta T_a$. For 1-2 ME, this is possible with plausible average heat flow, however for the more massive planets, thermal convection isn't possible and the cores will cool down to a sub-adiabatic conductive state.

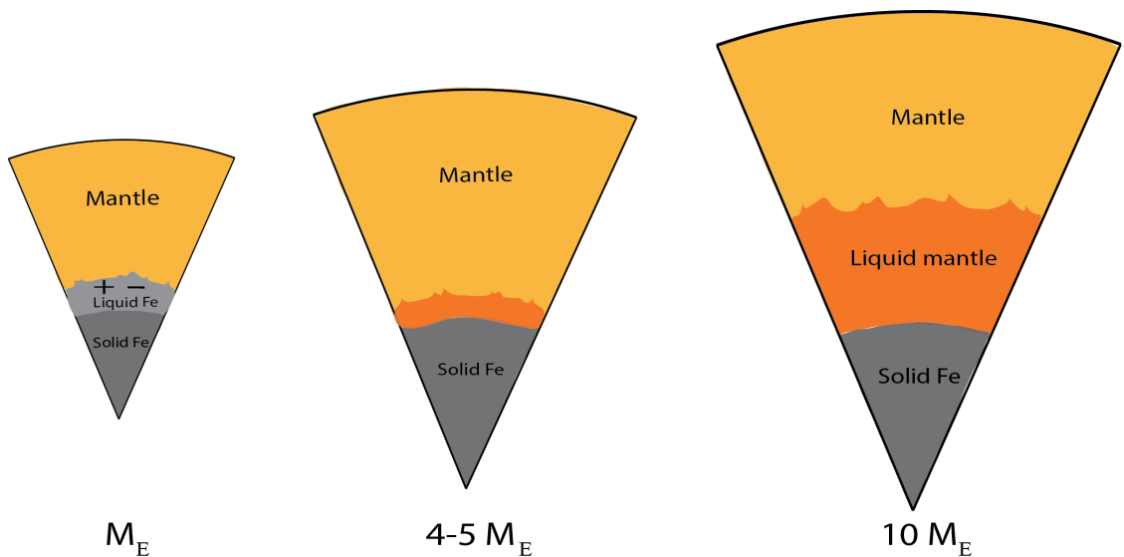

Figure 6 Internal models of terrestrial-like superEarths as a function of their planetary masses. Planets above 4 $M_E$ distinguish themselves from those with lower masses in that 1- they lack a liquid convecting Iron cores 2- they contain deep basel magma silicate oceans.

**Appendix**

To derive the density radial structure, one need to solve the following continuity euqations

$$\frac{d\rho}{dr} = -\frac{\rho^2(r)g(r)}{K_s(r)} \quad (1)$$

$$\frac{dg}{dr} = 4\pi G\rho(r) - \frac{2G\,m(r)}{r^3} \quad (2)$$

$$\frac{dP}{dr} = -\rho(r)g(r) \quad (3)$$

$$\frac{dm}{dr} = 4\pi r^2(r)\rho(r) \quad (4)$$

$$P = 3K_0 \frac{1-x}{x} \exp\left(\frac{3}{2}(K_0'-1)(1-x)\right) \quad (5)$$

Where $x = {\rho_0}/{\rho}$ , $K_s$ is the adiabatic bulk modulus while G is the gravitational constant.

For the melting depression of Iron alloys, one can thus use the Gibbs free energy rule to calculate the solidus line of the Iron-alloy as a function of impurity concentration. If the two phases, solid and liquid, are in thermal equilibrium, the Gibbs free energy, G, and the chemical potential across the phase boundary must be equal. Therefore,

$$\mu_{Fe}^l\,(P,T) + RT \sum \ln 1 - \chi_{i,l} = \mu_{Fe}^s\,(P,T) \quad (6)$$

Where $\mu_{Fe}^l$ and $\mu_{Fe}^s$ are the chemical potentials for the liquid and solid Fe phases respectively, whereas $\chi_{i,l}$ is the molar fraction of a given impurity in the liquid core denoted by i. By rewriting the equation above in terms of the enthalpy of fusion, H, and entropy S using $\Delta\mu = \Delta H - T\Delta S$

## Acknowledgment

The author acknowledges multiple insightful discussions with G.W. Collins. This material is based upon work supported by the Department of Energy National Nuclear Security Administration under Award Number DE-NA0003856, the University of Rochester, and the New York State Energy Research and Development Authority.